\documentclass[11pt,fleqn]{article}

\usepackage{amsmath} 
\usepackage{amsfonts}
\usepackage{graphicx}
\usepackage{dcolumn}
\usepackage{empheq}
\usepackage{upgreek}
\usepackage{latexsym,slashed}
\usepackage{amssymb,amsfonts}
\usepackage{geometry}
\usepackage{subfigure}
\usepackage{cancel}
\usepackage{bm} 
\usepackage{amsmath,mathrsfs} 

\usepackage[affil-it]{authblk}

\usepackage[parfill]{parskip}
\usepackage{booktabs} 
\usepackage{array} 
\usepackage{paralist} 
\usepackage{verbatim} 
\usepackage{subfig} 
\usepackage{multirow}
\usepackage{fancyhdr} 

\usepackage{sectsty}
\allsectionsfont{\sffamily\mdseries\upshape} 

\usepackage[nottoc,notlof,notlot]{tocbibind} 
\usepackage[titles,subfigure]{tocloft} 

\newcommand{\ack}[1]{\par\section*{Acknowledgement} #1}

\newcommand{\pacs}[1]{\smallskip\noindent{\sl PACS numbers: \hspace{0.3cm}#1}\par\bigskip\rm}
\newcommand{\address}[1]{\begin{center}\large #1\end{center}}
\begin{document}
\tolerance=5000

\title{Nonlinear bulk viscosity and the stability of accelerated expansion in FRW spacetime} 

\author{G. Acquaviva\footnote{acquavivag@unizulu.ac.za},
A. Beesham\footnote{beeshama@unizulu.ac.za}}
\date{}
\maketitle
\address{Department of Mathematical Sciences, University of Zululand, Private Bag X1001, Kwa-Dlangezwa 3886, South Africa}
\vspace{1cm}

\begin{abstract}
In the context of dark energy solutions, we consider a Friedmann-Robertson-Walker spacetime filled with a non-interacting mixture of dust and a viscous fluid, whose bulk viscosity is governed by the nonlinear model proposed in \cite{mm}.  Through a phase space analysis of the equivalent dynamical system, existence and stability of critical solutions are established and the respective scale factors are computed.  The results point towards the possibility of describing the current accelerated expansion of the Universe by means of the abovementioned nonlinear model for viscosity.
\end{abstract}


\section*{Introduction}

The discovery \cite{riess,perl} and confirmation of the present accelerated expansion of our Universe has opened up one of the main modern challenges in theoretical cosmology together with the early inflationary mechanism: the identification of the energy content responsible for such behaviour \cite{wein}.  In fact, assuming the  validity of Einstein's theory of gravitation at supergalactic scales ({\it i.e.}, disregarding here modifications in the geometric sector), an accelerated expansion can be obtained through a variety of energy-momentum tensor choices, among which we recall the introduction of scalar fields with suitable potentials, such as quintessence \cite{cald} or phantom fields \cite{cald2}.  However, these kinds of models are often subject to severe {\it fine tuning} problems and, even though the fine tuning argument in itself is not sufficient to rule out such possibilities, some kind of justification is expected to be provided from particle physics.  Another approach which allows one to obtain an accelerated expansion consists of modifying the simple barotropic equation of state (EoS) of the fluid in favour of more exotic forms, such as the Chaplygin gas model \cite{kame} and its generalizations \cite{bento}.  For a recent review of the efforts aimed at resolving this {\it dark energy} problem, see \cite{miao} and references therein.

In this work we focus on another kind of approach which, to some extent, tends to minimize the appeal to exotic forms of matter: the introduction of dissipative processes through the modeling of viscous effects in ordinary fluids.  Such an approach has already been shown to be able to produce accelerated growth of the scale factor in the context of Israel-Stewart's linear causal theory \cite{isra,coley},  its truncated causal form or the non-causal Eckart theory (see \cite{maart} and references therein).  On the other hand, some authors (such as \cite{avelino}) pointed out that the linear theory is not able to produce a cosmic evolution in accordance with observations.  The main problem with the linear Israel-Stewart theory is that it relies on the assumption of small deviations from thermodynamical equilibrium, an assumption that is not always expected to hold, in particular when accelerated expansion takes place and we are dealing with dark energy.  For this reason, several authors have tried to extend the theory of viscosity away from equilibrium \cite{novello,jou}, introducing for example nonlinear effects in the dynamics of the viscous pressure $\Pi$.  In the following we take into account the nonlinear model proposed in \cite{mm} and study the phase space of the theory in FRW spacetime in the presence of both dust and a generic viscous fluid governed by an equation of state (EoS) parameter $\gamma$.  Both local (critical points and their stability) and global features (characteristic trajectories and basins of attraction) are analyzed.  Such a model has been studied also in \cite{chim}, where it has been shown that the presence of a single viscous fluid can lead to a stable accelerated expansion.

\section{Relevant equations}
\label{rele}

Given the Friedmann-Robertson-Walker (FRW) metric
\begin{equation*}\label{met}
 ds^2 = -dt^2 + a^2(t)\left( dr^2 + r^2\, d\Omega^2 \right)\, ,
\end{equation*}
 the expansion scalar is defined as $\theta=3\dot{a}/a$.  We let  $8\pi G=1$. We consider the universe to be filled with a non-interacting mixture of dust, with energy density $\rho_d$ ($p_d=0$), and a viscous fluid with energy density $\rho_v$ and pressure $p_v=p_v(\rho_v)$. Then, from Einstein's field equations (EFEs),  we obtain the \textit{Raychaudhuri equation}
\begin{align}\label{ray}
 \dot{\theta} = -\frac{1}{3}\theta^2 - \frac{1}{2} \left(\rho_d + \rho_v + 3 p_v + 3\Pi \right)\, ,
\end{align}
and the \textit{Friedmann constraint equation}
\begin{equation}\label{fried}
  \rho_d + \rho_v - \frac{1}{3}\theta^2 = 0\,
\end{equation}
\textit{Conservation of energy-momentum} provides the other relevant evolution equations,
\begin{subequations}
\label{cons:tot}
\begin{empheq}[]{align}
  \dot{\rho_v} &= -\theta \left( \rho_v + p_v + \Pi \right) \label{cons:a} \\
  \dot{\rho_d} &= -\theta\, \rho_d\, . \label{cons:b}
\end{empheq}
\end{subequations}
We consider a barotropic EoS for the viscous fluid:
\begin{equation}\label{eos}
 p_v = (\gamma-1)\, \rho_v
\end{equation}

Using the EoS eq.(\ref{eos}) and the constraint eq.(\ref{fried}), eq.(\ref{ray}) and eq.(\ref{cons:a}) become
\begin{subequations}
\label{dyn:tot}
\begin{empheq}[left={}\empheqlbrace]{align}
  \dot{\theta} &= -\frac{1}{2}\theta^2 -\frac{3}{2} \Big[ (\gamma-1)\, \rho_v + \Pi \Big] \label{dyn:a} \\
  \dot{\rho_v} &= -\theta \left(\gamma\, \rho_v + \Pi \right) \label{dyn:b}
\end{empheq}
\end{subequations}

With regard to the evolution of the viscous pressure variable $\Pi$, we make reference to the nonlinear model proposed in \cite{mm}:
\begin{equation}\label{nonlin}
 \tau  \dot{\Pi} = -\zeta\,  \theta -\Pi \left(1+\Pi\,  \frac{\tau_*}{\zeta }\right)^{-1} -\frac{1}{2}\, \Pi\, \tau \left[\theta +\frac{\dot{\tau}}{\tau}-\frac{\dot{\zeta}}{\zeta}-\frac{\dot{T}}{T}\right]
\end{equation}
This equation\footnote{We refer to \cite{mm} for more details on the derivation and motivations.} has been derived by assuming a nonlinear relationship between the thermodynamic \textquotedblleft force\textquotedblright $\chi$ and the thermodynamic \textquotedblleft flux\textquotedblright\ $\Pi$ of the form
\begin{equation*}
\Pi = -\frac{\zeta\, \chi}{1+\tau_*\, \chi}
\end{equation*}
where $\zeta$ is the bulk viscosity and $\tau_*$ is the characteristic time for nonlinear effects (the linear Israel-Stewart theory is recovered for $\tau_*=0$).

The quantities $\tau$ and $T$ in eq.\eqref{nonlin} are the linear relaxational time and the local equilibrium temperature, respectively.  The quantity in round brackets is strictly positive in order to ensure positivity and non-divergence of the entropy production rate.  We will consider the following relations for the parameters involved (see \cite{mm}):
\begin{itemize}
 \item bulk viscosity: $\zeta=\zeta_0\ \theta$, with $\zeta_0>0$
 \item linear relaxational time:\ $\tau=\zeta/(\gamma\, v^2\, \rho_v)$
 \item nonlinear characteristic time: $\tau_*=k^2\, \tau$
 \item temperature (barotropic fluid): $T=T_0\, \rho^{(\gamma-1)/\gamma}$, from the integrability condition of the Gibbs relation
\end{itemize}
Here $v$ is the dissipative contribution to the speed of sound $V$, whose complete expression is $V^2=c_s^2 + v^2$.  From the condition $V^2\leq 1$ and from the definition of adiabatic sound speed $c_s^2\equiv \delta p/\delta \rho = \gamma-1$, we have the bound 
\begin{equation}
v^2\leq 2-\gamma\text{,\ \ with\ \ } 1\leq\gamma\leq2\label{sos}
\end{equation}
The explicit form of the evolution equation is then given by
\begin{equation}\label{pidot}
 \dot{\Pi} = -\gamma\, v^2\, \rho_v\, \theta - \frac{\gamma\, v^2}{\zeta_0} \frac{\Pi\, \rho_v}{\theta}\, \left( 1+\frac{k^2}{\gamma\, v^2}\frac{\Pi}{\rho_v} \right)^{-1} - \frac{1}{2}\, \Pi\, \left[ \theta - \left(\frac{2\gamma-1}{\gamma}\right) \frac{\dot{\rho}_v}{\rho_v} \right]
\end{equation}

\section{Dynamical system}

In order to reduce the dynamical equations to an autonomous system, we define the \textit{expansion-normalized variables} $\Omega=3\rho_v/\theta^2$ and $\tilde{\Pi}=3\Pi/\theta^2$, together with the new time variable $dt/d\tau=3/\theta$, whose associated derivative will be denoted by a prime.  The system eq.(\ref{dyn:tot}) in terms of the normalized variables is
\begin{subequations}
\label{dyn2:tot}
\begin{empheq}[left={}\empheqlbrace]{align}
 \frac{\theta'}{\theta} &= -\frac{3}{2} \left[ 1+ (\gamma-1)\, \Omega + \tilde{\Pi} \right] \label{dyn2:a} \\
 \frac{3\rho_v'}{\theta^2} &= -3 \left( \gamma\, \Omega + \tilde{\Pi} \right)\ . \label{dyn2:b}
\end{empheq}
\end{subequations}
From the definition of $\Omega$, we obtain
\begin{equation}
 \Omega' = \frac{3\, \rho_v'}{\theta^2} - 2\, \Omega\, \frac{\theta'}{\theta}\, .
\end{equation}
Substituting eqs.(\ref{dyn2:tot}) in this last equation, we get the evolution equation for $\Omega$
\begin{equation}\label{omegaprime}
 \Omega' = 3\left( \Omega-1\right) \left[ (\gamma-1)\, \Omega + \tilde{\Pi} \right]
\end{equation}

We now introduce the evolution equation for $\tilde{\Pi}$.  Differentiating $\tilde{\Pi}$ with respect to $\tau$ and making use of eq.(\ref{dyn2:a}), we get
\begin{align}
 \tilde{\Pi}' = &-3\gamma\, v^2\, \Omega \left[ 1+ \frac{\tilde{\Pi} }{3\, \zeta_0}\, \left( 1+\frac{k^2}{\gamma\, v^2}\frac{\tilde{\Pi}}{\Omega} \right)^{-1} \right] - 3\, \frac{\tilde{\Pi}^2}{\Omega}\, \left[ \frac{2\gamma-1}{2\gamma}-\Omega \right] +\nonumber\\
 &- 3\, \tilde{\Pi} (\gamma-1)\, (1- \Omega) \label{piprime}
\end{align}
Thus the dynamical system is described by eqs.(\ref{omegaprime}) and (\ref{piprime}).  The search for critical points amounts to looking for values $\{ \Omega_c\, ,\, \tilde{\Pi}_c \}$ which solve the system $\Omega'=\tilde{\Pi}'=0$.

\section{Phase space analysis}

First of all, we stress that positivity of entropy production rate requires a restriction of the phase space to the region
\begin{equation}\label{posentr}
 \tilde{\Pi} > -\frac{\gamma\, v^2}{k^2}\, \Omega
\end{equation}
Note that if $k^2\gg v^2$, this condition narrows the possible negative values of $\tilde{\Pi}$ toward zero.  For finite $k$, in the limit $v\rightarrow 0$, \textit{only positive values of bulk pressure are allowed}.  On the contrary, for $k^2\ll v^2$ the bounds on bulk pressure are less restrictive.  A good tradeoff would be to consider $k^2 \lesssim v^2$, which, through the fact that $v^2\leq 2-\gamma$ and the definition $\tau_*=k^2\, \tau$, means that the characteristic time for nonlinear effects $\tau_*$ does not exceed the characteristic time for linear relaxational effects $\tau$.

Important quantities in the characterization of critical points will be the deceleration parameter $q=-1-\theta'/\theta$ and the effective EoS parameter $\ \gamma_{eff}=-2\theta'/3\theta$, which are given respectively by
\begin{align}
 q &= \frac{1}{2} \left[ 1 + 3 (\gamma-1)\, \Omega +3 \tilde{\Pi} \right]\label{acc}\\
 \gamma_{eff} &=1+(\gamma-1)\, \Omega + \tilde{\Pi}\label{effeos}
\end{align}
The region of the phase space where accelerated expansion occurs is found by imposing $q<0$ in eq.(\ref{acc}):
\begin{equation*}
 \tilde{\Pi} < -\frac{1}{3} - (\gamma-1)\Omega
\end{equation*}
In this region the effective EoS parameter $\gamma_{eff}<2/3$.  Comparing eq.(\ref{posentr}) and eq.(\ref{acc}) for $q<0$, one finds that accelerated expansion is possible in the physical phase space only if
\begin{equation}\label{accecond}
 \frac{v^2}{k^2}>\frac{1+3\, \Omega\, (\gamma -1)}{3\, \gamma\, \Omega}
\end{equation}

Considering eq.(\ref{omegaprime}), letting $\Omega'=0$, we identify two conditions: 
\begin{subequations}
\label{subs:tot}
\begin{empheq}[left={}\empheqlbrace]{align}
 &\Omega_c = 1 \label{subs:a} \\
 &(\gamma-1)\, \Omega_c + \tilde{\Pi}_c=0 \label{subs:b}
\end{empheq}
\end{subequations}
In order to locate the critical points one has to substitute these conditions into the  equation $\tilde{\Pi}'=0$.
We will carry out the analysis specifying the character of the viscous fluid through the choice of $\gamma$.  Moreover, given the considerations above, from now on we will set $0<k^2=v^2\leq2-\gamma$.  We note that this choice excludes the case of stiff matter ($\gamma=2$) from the analysis, because the bound on the dissipative speed of sound would give $v^2=0$.

\subsection{Dust: $\gamma=1$}

The first condition $\Omega_c=1$ identifies an invariant set, in which equation $\tilde{\Pi}'=0$ is given by
\begin{equation*}
 \frac{3}{2}\, \tilde{\Pi}_c^3+\frac{3}{2}\, \tilde{\Pi}_c^2-\frac{v^2}{\zeta_0}\left(1+3\, \zeta_0 \right)\, \tilde{\Pi}_c -3 v^2 = 0
\end{equation*}
In the range of parameters considered, this cubic equation has 3 real roots and only one of them is positive.  Furthermore, it can be shown that the most negative root lies always in the region of negative entropy production rate, so that we are concerned only with the other two roots $\tilde{\Pi}^+$ and $\tilde{\Pi}^-$, where $\pm$ specifies their sign.  The system has thus 2 critical points in the invariant set $\Omega_c=1$ and we call them $P_d^+$ and $P_d^-$.  Given the dynamical system in the general form
\begin{subequations}
\label{dynimp:tot}
\begin{empheq}[left={}\empheqlbrace]{align}
 &\Omega' = f\left(\Omega,\tilde{\Pi}\right) \label{dynimp:a} \\
 &\tilde{\Pi}'= g\left(\Omega,\tilde{\Pi}\right) \label{dynimp:b}
\end{empheq}
\end{subequations}
the stability of the critical points is analyzed through the eigenvalues of the matrix
\begin{equation}\label{matrix}
 A = \begin{pmatrix}
   \frac{\partial f}{\partial \Omega} & \frac{\partial f}{\partial \tilde{\Pi}} \\
   \frac{\partial g}{\partial \Omega} & \frac{\partial g}{\partial \tilde{\Pi}}
  \end{pmatrix}_{\vert{P_d^{\pm}}}
\end{equation}
If both eigenvalues have negative (respectively positive) real part then the point is a source (respectively a sink).  A saddle point is found if the real parts of the eigenvalues have opposite signs.  We find that $\partial_{\tilde{\Pi}} f\, {\vert_{P_d^{\pm}}} \equiv 0$, so that the eigenvalues are given by
\begin{align}
 \lambda_1 & \equiv \frac{\partial f}{\partial \Omega}\vert_{P_d^{\pm}} = 3 \tilde{\Pi}^{\pm}\\
 \lambda_2 & \equiv \frac{\partial g}{\partial \tilde{\Pi}}\vert_{P_d^{\pm}} = 3 \tilde{\Pi}^{\pm} - \frac{v^2}{\zeta_0\, (1+\tilde{\Pi}^{\pm})^2}
\end{align}
For the point $P_d^+$, given that $\tilde{\Pi}^+>0$, the sign of both eigenvalues is always positive, so it is a source point.  As regards $P_d^-$, instead the sign of the eigenvalues is negative, thus identifying a stable sink.

Considering now the condition eq.\eqref{subs:b}, which in the case of dust simplifies to $\tilde{\Pi}_c=0$, it has to be noted that it is not an invariant set, but it represents a line of points in the phase space where the flow is \textquotedblleft momentarily at rest\textquotedblright\ in the $\Omega$ direction (a set of turning points).  We can find a critical point on this line by imposing also the requirement $\tilde{\Pi}'=0$, whose solution is $\Omega_c=0$.  However, the stability analysis for the point $P_d^0=\{ 0,0 \}$ is hampered by its location, {\it i.e.}, it lies on the entropy production rate divergence line, where the system is not well defined.  In order to have an idea of the behaviour around this point, we resort to numerical plots.
\begin{figure}[h!]
 \begin{center}
 \subfigure[\hspace{-0.6cm}]{\includegraphics[width=6.8cm]{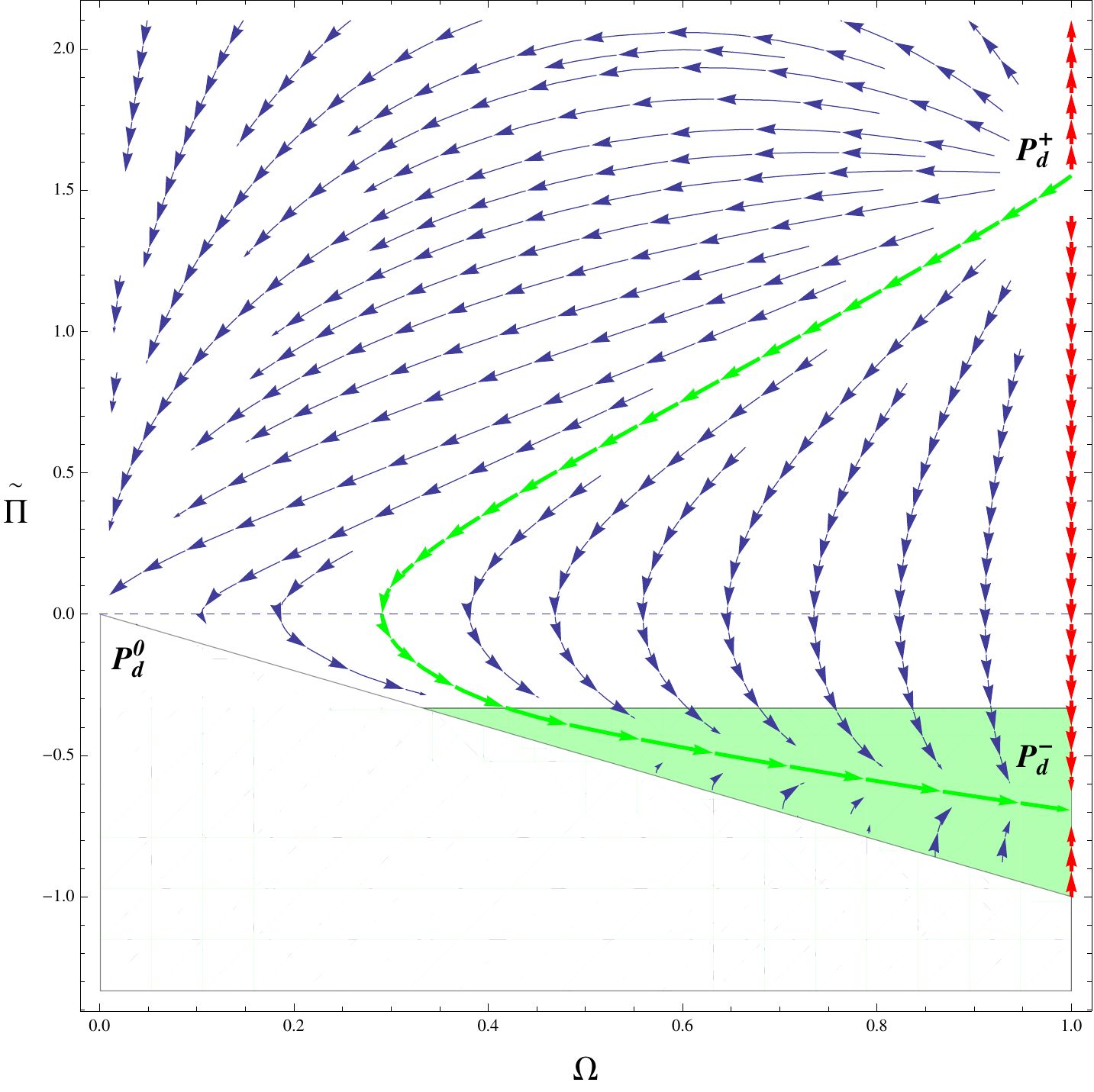}}\hspace{0.8cm}
 \subfigure[\hspace{-0.6cm}]{\includegraphics[width=6.8cm]{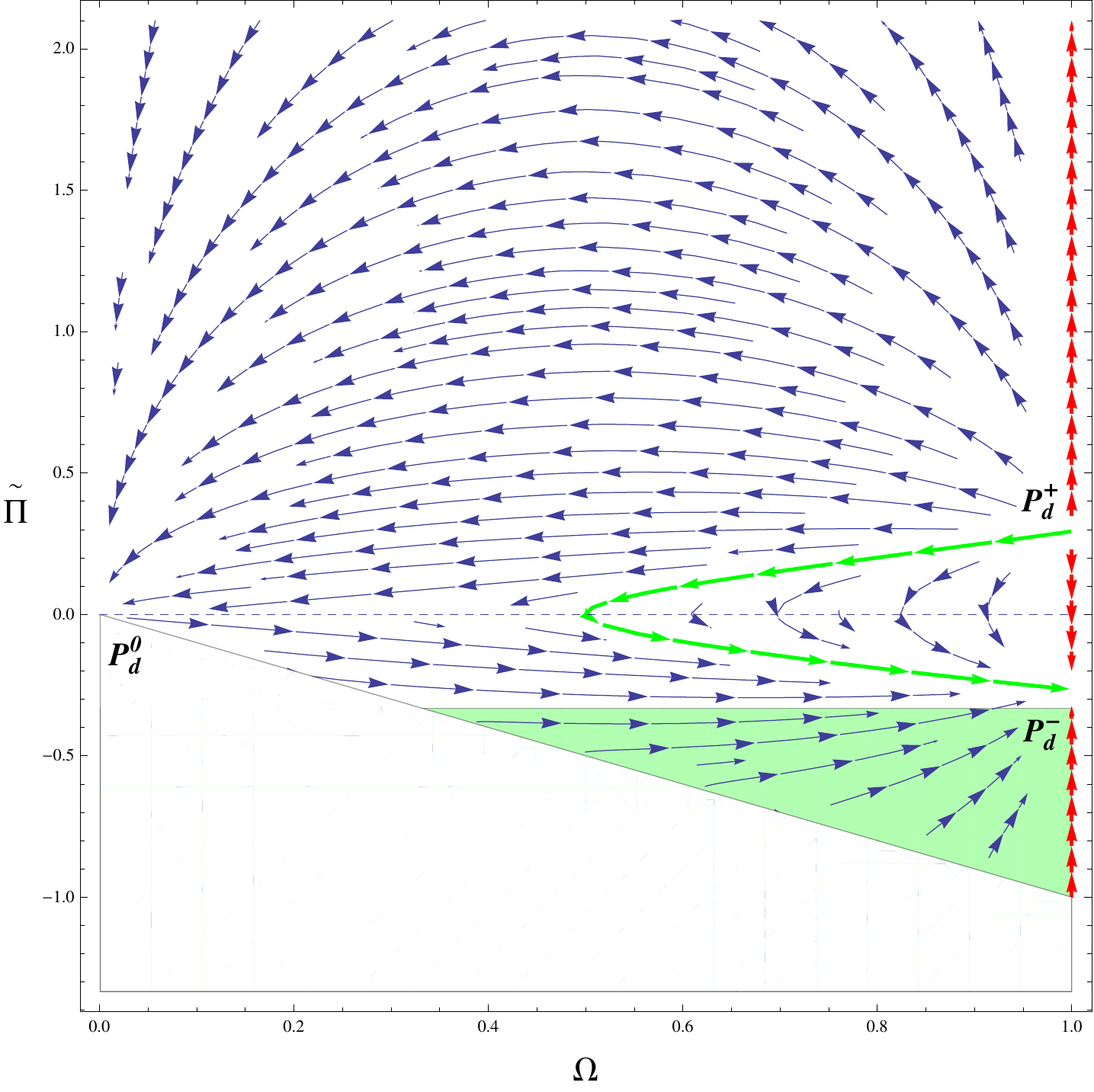}}
 \caption{\label{dyn_dust} Phase plane evolution of the system with $\gamma=1$ (dust) and $\zeta_0=1$.  (a) $v^2=k^2=1$, (b) $v^2=k^2=1/25$.  \textit{Red flow:} invariant set $\Omega=1$.  \textit{White region:} negativity of entropy production rate. \textit{Green region:} accelelerated expansion.    \textit{Dashed plot:} the line $\tilde{\Pi}=0$.}
\end{center}
\end{figure}
The global situation is depicted in Fig.~\ref{dyn_dust}: the green trajectory is a representative flow going from $P_d^+$ towards $P_d^-$, while the red trajectory is the invariant set $\Omega=1$.  The white region has been cut out from the analysis because it corresponds to models with negative entropy production rate, whereas on its boundary, the rate diverges.  Note that this boundary is {\it repulsive}, in the sense that no trajectory in its neighbourhood is attracted towards it: this feature keeps the models safe from divergences in entropy production rate.  In the green region $q<0$, so the expansion is accelerated.  As can be seen, the point $P_d^0$ turns out to be a saddle, while $P_d^-$ is the global attractor for every choice of initial conditions in the physical region and it describes a cosmological expanding model dominated by viscous matter.  In the portrait on the left, the choice of the parameters leads to an accelerated expansion, while in the portrait on the right the stable point lies in the deceleration region.  In Table \ref{stab_dust} a summary of the stability analysis is presented.
\vspace{0.5cm}
\setlength{\extrarowheight}{10pt}
\begin{table}[h!]
\centering
  \begin{tabular}{ | c || c | c | c |}
    \hline
    {\bf Point:} &  $P_d^0$ & $P_d^+$ & $P_d^-$\\[0.2cm] \hline
    {\bf Character:} & saddle & source & sink \\[0.2cm] \hline
  \end{tabular}
\caption{Stability of critical points for $\gamma=1$.}
\label{stab_dust}
\end{table}

\subsection{Radiation: $\gamma=4/3$}

Just like before, we start imposing the condition eq.\eqref{subs:a} in equation $\tilde{\Pi}'=0$, obtaining the third order equation
 \begin{equation*}
  \frac{27}{32}\, \tilde{\Pi}_c^3+\frac{9}{8}\, \tilde{\Pi}_c^2-\frac{v^2}{\zeta_0}\left(\frac{4}{3}+3\, \zeta_0 \right)\, \tilde{\Pi}_c -4 v^2 = 0
 \end{equation*}
Among the three real roots, we again retain only the two that lie in the physical phase space.  Hence we find two critical points, $P_r^+=\{ 1, \tilde{\Pi}_c^+ \}$ and $P_r^-=\{ 1, \tilde{\Pi}_c^- \}$, where $\tilde{\Pi}_c^+$ (resp. $\tilde{\Pi}_c^-$) is the positive (resp. negative) root.  

Imposing in $\tilde{\Pi}'=0$ the second condition eq.(\ref{subs:b}), which now reads $\tilde{\Pi}_c=-\Omega_c/3$, we find the critical point $P_r^0=\{0,0\}$ and another critical point
\begin{equation}
 P_r^* = \left\{ \frac{27\, \zeta_0}{4\, v^2}\, \left(v^2-\frac{1}{32}\right)\ ,\ -\frac{9\, \zeta_0}{4\, v^2}\, \left(v^2-\frac{1}{32}\right) \right\}\label{pstar}
\end{equation}
whose presence and influence in the physical phase space is subject to the condition $\Omega_c^*\leq 1$, that is
\begin{subequations}
\label{pstar_cond:tot}
\begin{empheq}[left={}\empheqlbrace]{align}
 \forall\, v>0\ \ \ &\text{if}\ \ \ 0<\zeta_0<\bar{\zeta}_0 \label{pstar_cond:a}\\
 0<v\leq \bar{v}\ \ \ &\text{if}\ \ \ \zeta_0> \bar{\zeta}_0\label{pstar_cond:b}
\end{empheq}\vspace{0.2cm}
\begin{equation*}
 \text{where}\ \ \ \bar{v}\equiv\sqrt{\frac{\zeta_0}{32\, \left( \zeta_0-\bar{\zeta}_0 \right)}}\ \ \ \text{and}\ \ \ \bar{\zeta}_0\equiv4/27
\end{equation*}
\end{subequations}

The dynamics and stability properties in the phase space are similar to the case of dust only when $P_r^*$ is not present, but an important difference arises when considering the stability of the critical point $P_r^-$ in conjunction with the appearance of $P_r^*$.

\begin{figure}[h!!!]
 \begin{center}
 \subfigure[\hspace{-0.6cm}]{\includegraphics[width=7cm]{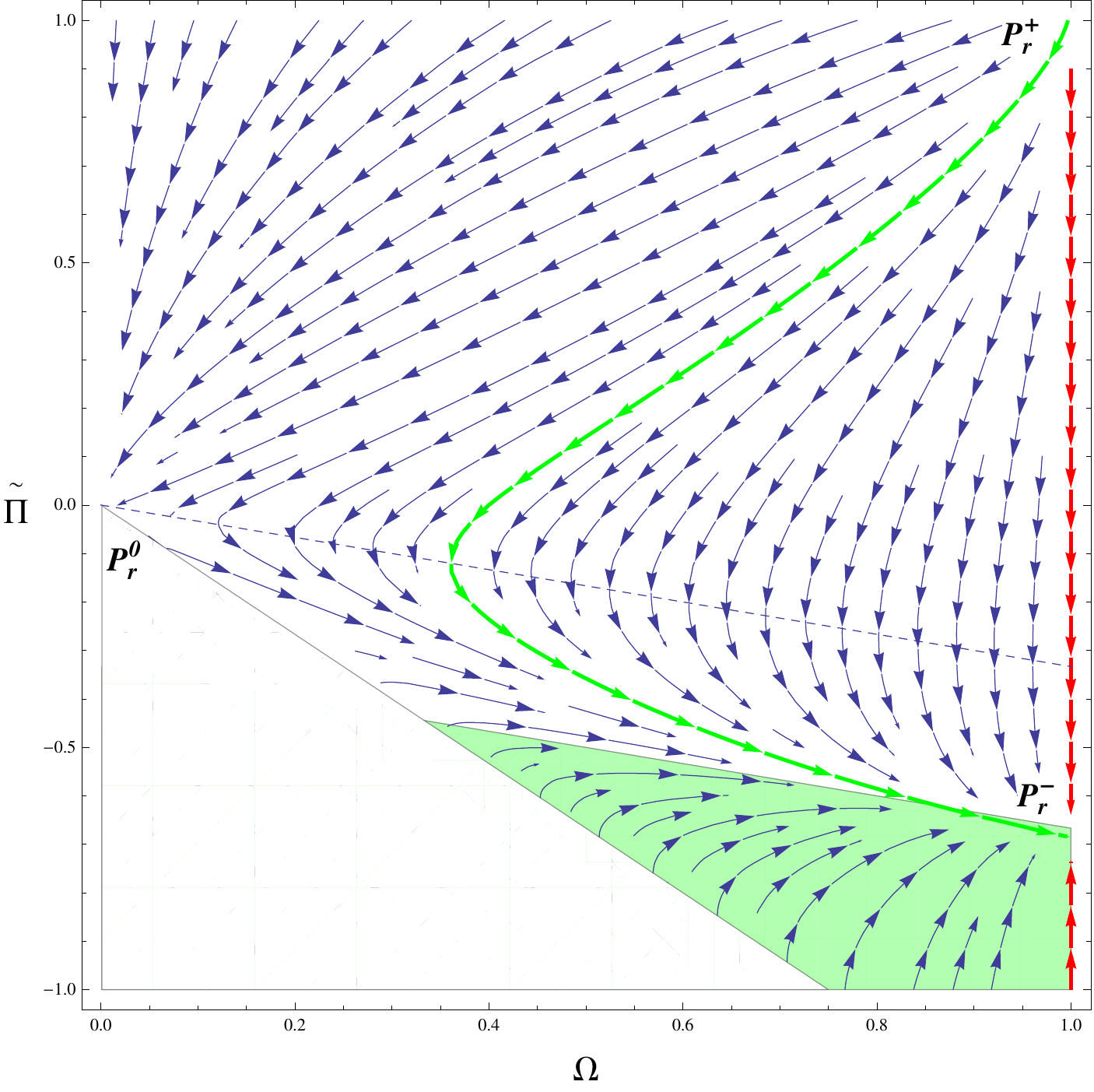}}\hspace{0.8cm}
 \subfigure[\hspace{-0.6cm}]{\includegraphics[width=7cm]{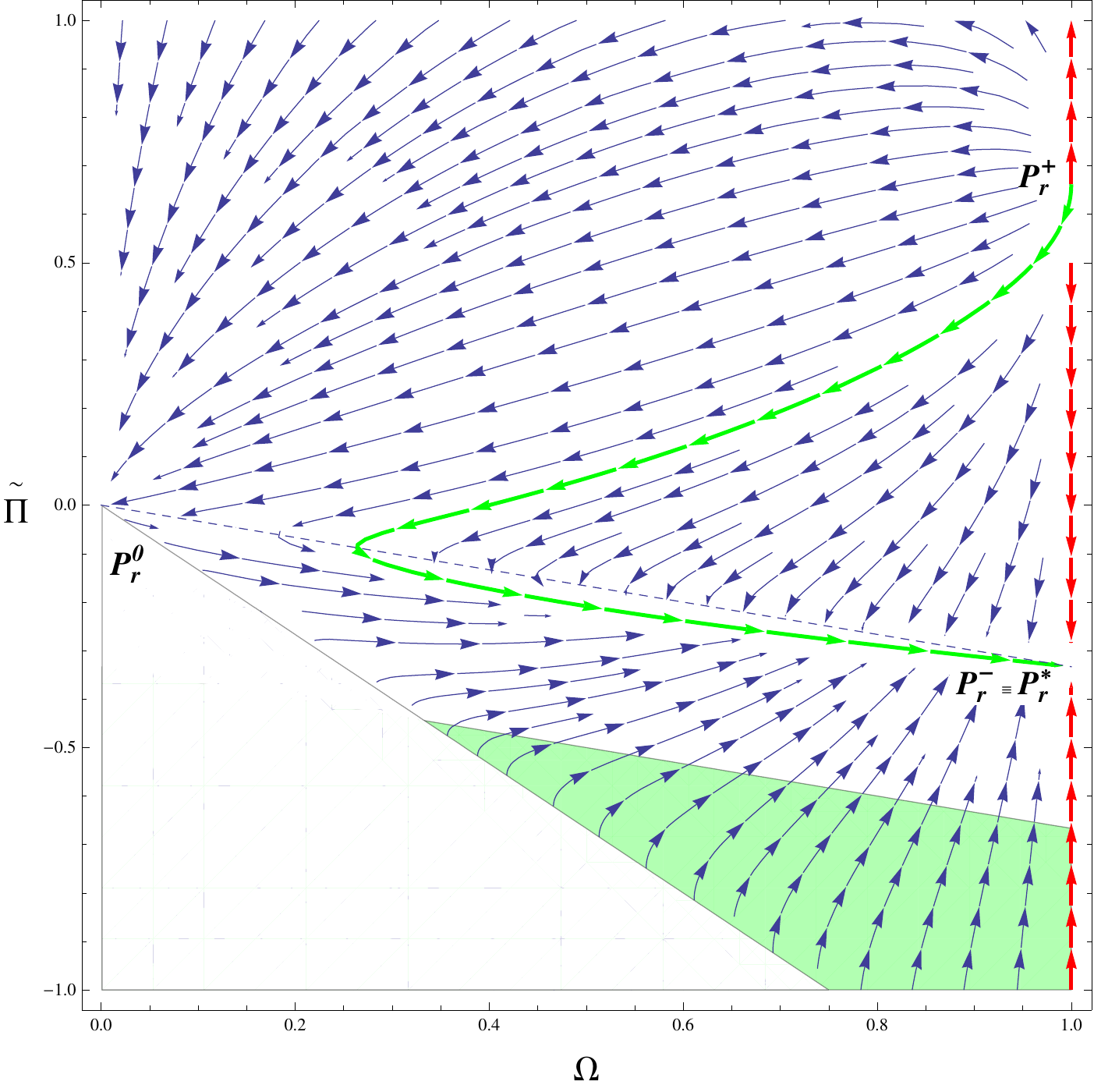}}
 \subfigure[\hspace{-0.6cm}]{\includegraphics[width=7cm]{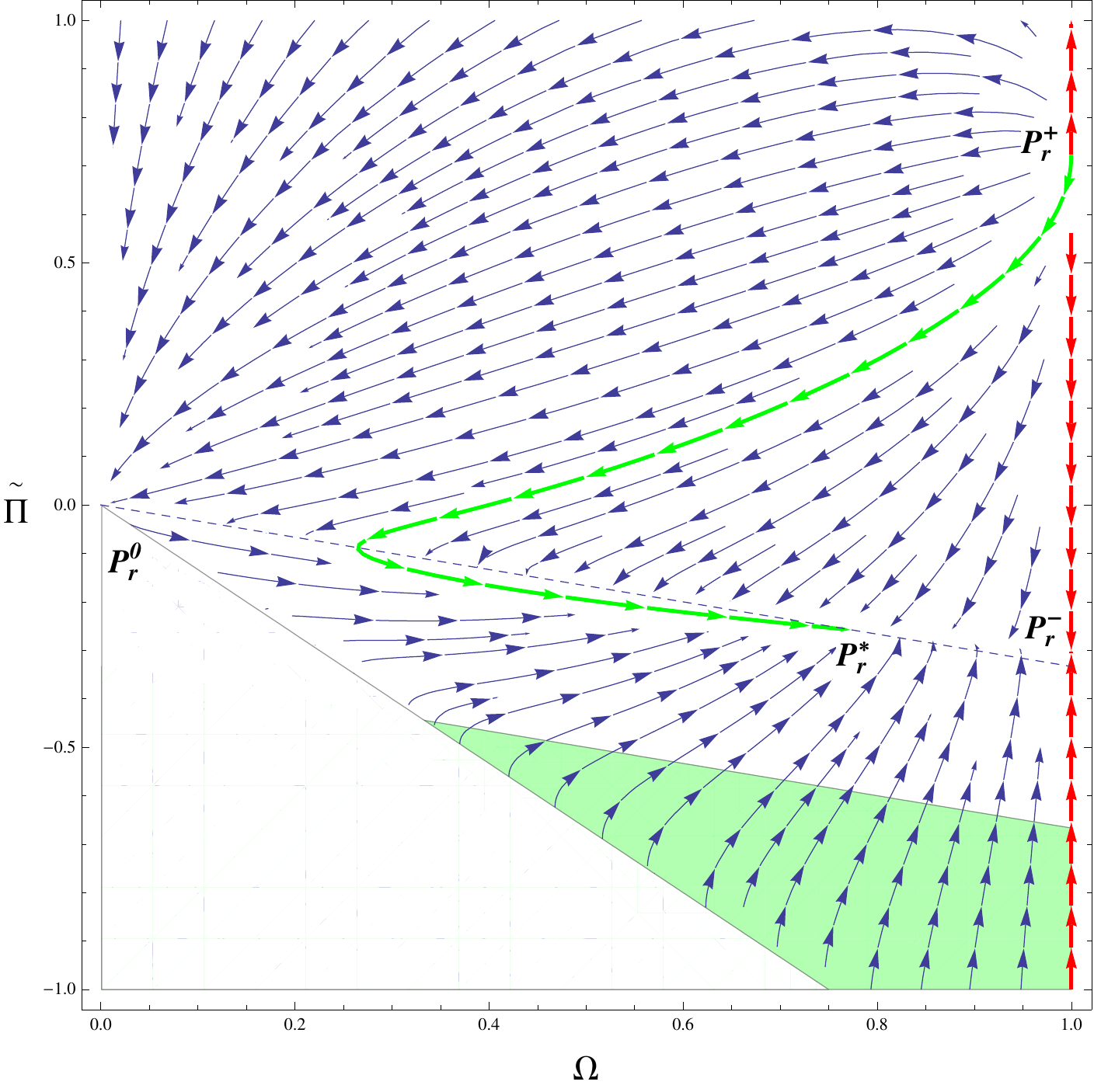}}
 \caption{\label{dyn_rad} Phase plane evolution of the system with $\gamma=4/3$ (radiation) and: (a) $v=\sqrt{2/3}$, $\zeta_0=1$,  (b) $v=\bar{v}$, $\zeta_0=\bar{\zeta_0}+1/10$, (c) $v=\bar{v}-1/10$, $\zeta_0=\bar{\zeta}_0+1/25$.  \textit{Red flow:} invariant set $\Omega=1$.  \textit{White region:} negativity of entropy production rate. \textit{Green region:} accelelerated expansion.  \textit{Dashed plot:} the line $\tilde{\Pi}=-\Omega/3$.}
\end{center}
\end{figure}

Let us analyze the stability properties in more detail.  As in the previous case, we defer the qualitative stability analysis of $P_r^0$ to the numerical plot Fig.~\ref{dyn_rad}: this critical point is a saddle.  For the points $P_r^{\pm}$, the eigenvalues of the stability matrix eq.(\ref{matrix}) are
\begin{equation}
 1 + 3\, \tilde{\Pi}_c^{\pm}\ \ \ ,\ \ \ \frac{9\, \tilde{\Pi}_c^{\pm}}{4}-\frac{64\, v^2}{3\, \zeta_0\, (4+3\, \tilde{\Pi}_c^{\pm})^2}
\end{equation}
The critical point $P_r^+$ is always a source.  Critical point $P_r^-$ turns out to be a sink for $\tilde{\Pi}_c^-<-1/3$, which occurs outside the domain of parameters given by (\ref{pstar_cond:tot}), \textit{i.e.}, when $P_r^*$ is absent; if instead the parameters satisfy conditions (\ref{pstar_cond:tot}), then $P_r^-$ is a saddle and the \textquotedblleft newly born\textquotedblright\ critical point $P_r^*$ turns out to be a sink, having eigenvalues
\begin{equation*}
 -\frac{4\, \left(1+64\, v^2\right)\pm\sqrt{16\, v^2\, \left(5-64\, v^2\right)^2+81\, \left(1-32\, v^2\right)^2\, \zeta_0}}{32}
\end{equation*}
which are both real and negative in the ranges given by (\ref{pstar_cond:tot}).  A summary of the stability properties is given in Table \ref{stab_rad}.

\setlength{\extrarowheight}{10pt}
\begin{table}[h!!]
\centering
\begin{tabular}{|c||c||c|c|c|c|}
\hline
{\bf Point:} & \ & $P_r^0$ & $P_r^+$ & $P_r^-$ & $P_r^*$  \\[0.3cm]\hline\hline
 \multirow{2}{*}{\vspace{-0.6cm}{\bf Character:}} & {\it if (\ref{pstar_cond:tot}) hold} & saddle & source & saddle & sink \\[0.3cm] \cline{2-6}
& {\it if (\ref{pstar_cond:tot}) do not hold} & saddle & source & sink & (n/a)\\[0.3cm] \hline
\end{tabular}
\caption{Stability of critical points for $\gamma=4/3$.}
\label{stab_rad}
\end{table}

The dynamical system is plotted in Fig.~\ref{dyn_rad} for three representative sets of values of the parameters.  In Fig.~\ref{dyn_rad}(a) the global attractor $P_r^-$ represents an accelerated expanding model dominated by viscous radiation.  In Fig.~\ref{dyn_rad}(b) the point $P_r^-\equiv P_r^*$ is again the global attractor, but the model is decelerating.  Fig.~\ref{dyn_rad}(c) displays the peculiarity with respect to the dust model: the point $P_r^-$ is now an unstable saddle and the point $P_r^*$, located along the (dashed) line $\tilde{\Pi}=-\Omega/3$, is the new global attractor.  This point represents a decelerated expanding model with contributions coming both from viscous radiation and non-viscous dust.
\vspace{1cm}

\section{Scale factors}

\begin{figure}[h!!!]
 \begin{center}
 \subfigure[\hspace{-0.6cm}]{\includegraphics[width=6.5cm]{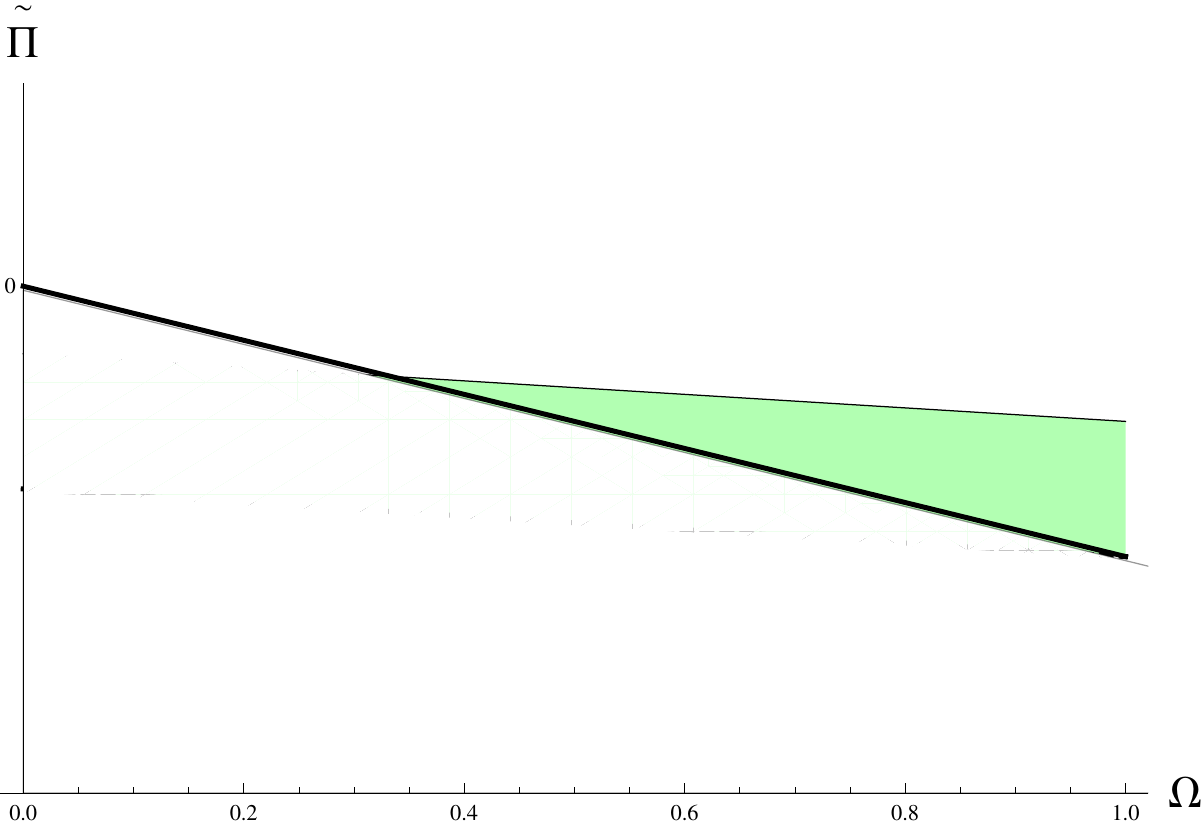}}\hspace{0.8cm}
 \subfigure[\hspace{-0.6cm}]{\includegraphics[width=6.5cm]{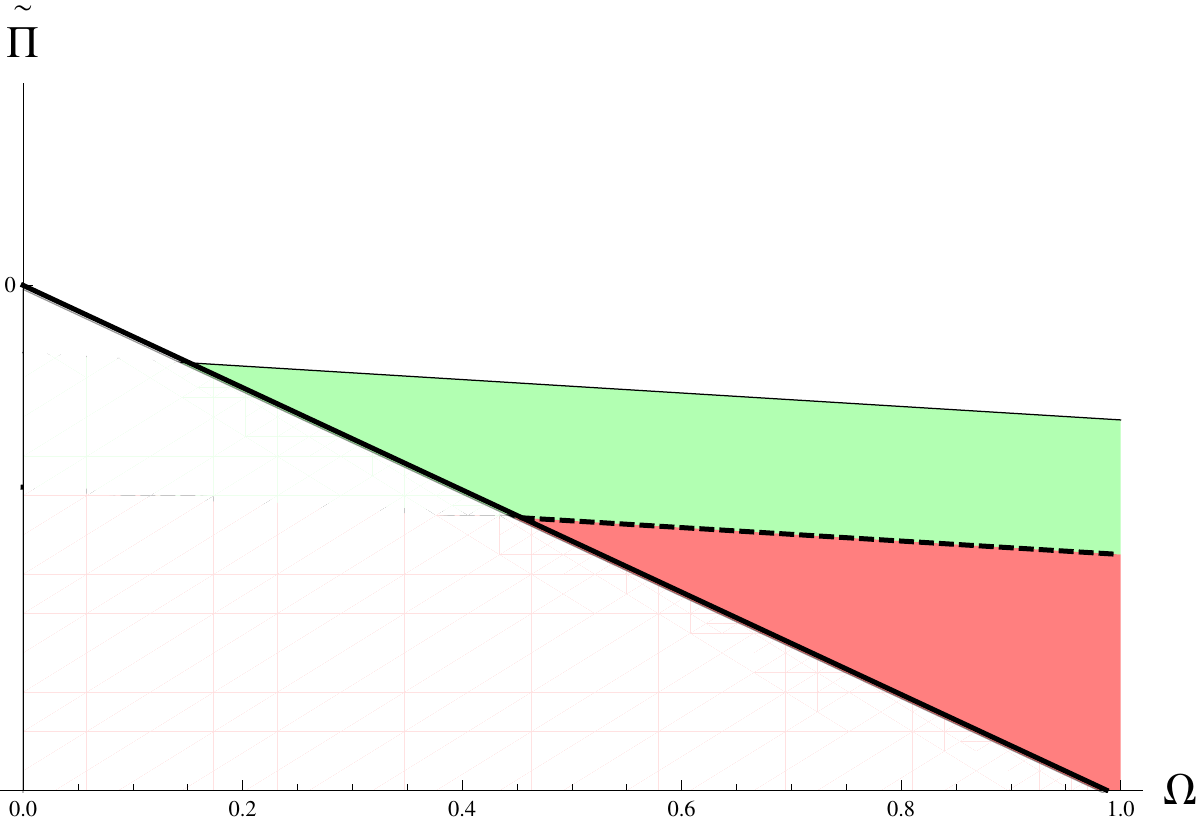}}
 \caption{\label{expo} Qualitative phase space portrait for (a) $v^2=k^2$ and (b) $v^2>k^2$.  The physical phase space is above the {\it thick line}.  {\it Green region:} accelerated expansion.  {\it Dashed line:} exponential behaviour.  {\it Red region:} contracting models.}
\end{center}
\end{figure}

Some general considerations on the form of the scale factors  corresponding to the  critical points can be made by integrating the cosmic-time version of eq.\eqref{dyn2:a}, {\it i.e.}
\begin{equation}\label{expan}
 \dot{\theta} = -\frac{1}{2}\, \left[ 1+(\gamma-1)\, \Omega + \tilde{\Pi} \right]\, \theta\, ,
\end{equation}
and then solving $\theta\equiv3\, \dot{a}/a$ for $a(t)$.  Power law scale factors are obtained whenever $1+(\gamma-1)\, \Omega + \tilde{\Pi} \neq 0$, because then $\dot{\theta}\neq0$.  In this case the result corresponding to a generic critical point is
\begin{equation}\label{scale}
 a(t) = a_0\, \left( t - t_0 \right)^{\frac{2}{3\left[ 1+(\gamma-1)\, \Omega_c + \tilde{\Pi}_c \right]}}
\end{equation}
It is easy to show that the choice $k^2=v^2$ made in the previous sections prevents the occurence of models with exponential scale factors.  In order for the exponentially expanding models (identified by the condition $1+(\gamma-1)\, \Omega + \tilde{\Pi} = 0$ in eq.\eqref{expan}) to be present in the physical region of phase space (bounded by eq.\eqref{posentr}), the following inequality must hold:
\begin{equation*}
 (1-\gamma)\, \Omega -1\, > -\frac{\gamma\, v^2}{k^2}\, \Omega\, ,
\end{equation*}
For $v^2=k^2$ the inequality does not hold in the physical phase space.  On the other hand, only if we let $v^2 > k^2$, can the inequality  be satisfied (consistently with the analysis leading to eq.(39) in \cite{mm}), specifically in the phase-space region
\begin{equation*}
 \left(1-\gamma \left(1-\frac{v^2}{k^2}\right)\right)^{-1}<\Omega \leq 1\ \ \ \wedge \ \ \ -\gamma <\Pi <\frac{2}{3} -\gamma
\end{equation*}
Moreover in this case, from eq.\eqref{scale}, it is clear that for $1+(\gamma-1)\, \Omega + \tilde{\Pi} < 0$, {\it i.e.}, below the \textquotedblleft exponential behaviour\textquotedblright\, line, we can find cosmological contracting models.  In Fig.~\ref{expo} a qualitative representation of the situation is plotted.

Finally, if we let instead $v^2<k^2$, the possibility of having accelerated expansion will narrow down, disappearing from the physical phase space when $v^2<\frac{\gamma-2/3}{\gamma}\, k^2$.

With regard to the critical points analyzed in the previous section, in Table \ref{scalef} are listed the respective polynomial scale factors.  The presence of the viscous pressure affects the power law by slowing down the dynamics in $P^+$ and accelerating it in $P^-$ with respect to the pure non-viscous fluid case.  The behaviour in the points $P^0$ and $P^*$ (actually on the whole line $(\gamma-1)\, \Omega + \tilde{\Pi}=0$, dashed in Figs. \ref{dyn_dust} and \ref{dyn_rad}) is always dust-like.

\setlength{\extrarowheight}{10pt}
\begin{table}[h!!]
\centering
\begin{tabular}{|c||c|c|c|c|}
\hline
{\bf Point:} & $P^0$ & $P^+$ & $P^-$ & $P^*$  \\[0.3cm]\hline\hline
{\bf $\gamma=1$} & $a_0\, (t-t_0)^{2/3}$ & $a_0\, (t-t_0)^{\frac{2}{3(1+\tilde{\Pi}^+_c)}}$ & $a_0\, (t-t_0)^{\frac{2}{3(1+\tilde{\Pi}^-_c)}}$ & (n/a) \\[0.3cm]\hline
{\bf $\gamma=4/3$} & $a_0\, (t-t_0)^{2/3}$ & $a_0\, (t-t_0)^{\frac{2}{4/3+\tilde{\Pi}^+_c}}$ & $a_0\, (t-t_0)^{\frac{2}{4/3+\tilde{\Pi}^-_c}}$ & $a_0\, (t-t_0)^{2/3}$\\[0.3cm] \hline
\end{tabular}
\caption{Scale factors corresponding to the critical points for $1+(\gamma-1)\, \Omega_c + \tilde{\Pi}_c \neq 0$ (polynomial behaviour).}
\label{scalef}
\end{table}

\section{Discussion}

From the analysis carried out, it is clear that stable solutions exist for non-interacting two-fluids models in the presence of nonlinear bulk viscous effects and that some of these solutions display an accelerated expansion.  If the viscous fluid is of dust type it will ultimately dominate  the non-viscous component; if the viscous fluid is of radiation type, its dominance upon the non-viscous dust depends on the parameters involved.

From observations, the current value of the deceleration parameter is around $q_0\simeq-0.5$ \cite{ama,bamba}.  For both fluid choices, the model identified by the stable critical point $P_i^-$ ($i=d,r$) is able to sustain an accelerated expansion with $-1<q<0$ whenever $\gamma-\frac{2}{3}<|\tilde{\Pi}_c|<\gamma$, having $\Omega_c=1$.  The accelerated future attractor model in this case has an effective EoS parameter in the range $0<\gamma_{eff}\vert_c<2/3$\, .

We highlight the fact that, in the case of viscous radiation, a trajectory in the phase space starting from a neighbourhood of $P_r^+$ with appropriate initial conditions can pass through the following stages: {\it i)} a radiation-dominated era (source $P_r^+$), {\it ii)} a matter-dominated transient era (saddle $P_r^0$), where structure formation can occur and {\it iii)} a final, everlasting era characterized by accelerated (either polynomial or exponential) expansion for a non-zero-measure set of parameters (sink $P_r^-$). This is in contrast to the findings of \cite{avelino} in which Eckart theory was used, and where it was found that radiation and matter solutions were not stable solutions. 

The kind of evolution that we have found shares similarities with the current accepted model for cosmic evolution.  Apart from the similarities, we stress that, at this level, the model {\it a)} does not include an explanation for primordial inflation and {\it b)} obviously shares with $\Lambda$CDM the ignorance about the future behaviour of the Universe, in the sense that in both models the accelerating phase (if present) lasts forever; the everlasting acceleration, in turn, is impossible to disprove without knowing {\it a priori} the exact matter-energy content.  In fact, it has to be kept in mind that such type of analysis is always phenomenological in nature.

\ack{The authors are thankful to the Astrophysics and Cosmology Research Unit (ACRU) at the University of KwaZulu-Natal for the kind hospitality, in particular Sunil Maharaj.  We also thank Rituparno Goswami and Radouane Gannouji for useful discussions.  GA thanks the University of Zululand for the award of a Postdoctoral Fellowship.}


\begin{thebibliography}{99}

\bibitem{riess} Riess A. G., \textit{et al.}, A. J. 116, 3 (1998): 1009

\bibitem{perl} Perlmutter, S. {\it et al} Ap. J. 517 (1999): 565 

\bibitem{wein} Weinberg D. H. {\it et al.}, Phys. Rep. 530 (2013): 87-255

\bibitem{cald} Caldwell R. R. {\it et al.}, Phys. Rev. Lett. 80, 8 (1998): 1582

\bibitem{cald2} Caldwell R. R. {\it et al.}, Phys. Rev. Lett. 91, 7 (2003): 071301

\bibitem{kame} Kamenshchik A. {\it et al.}, Phys. Lett. B  511, 2 (2001): 265-268

\bibitem{bento} Bento M. C. {\it et al.}, Phys. Rev. D 66, 4 (2002): 043507

\bibitem{miao} Li M. {\it et al.}, Front. Phys. 8, 6 (2013): 828-846

\bibitem{isra} Israel W. and Stewart J. M., Ann. Phys. 118, 2 (1979): 341-372

\bibitem{coley} Coley A. A. and Van den Hoogen R. J., Class. Quant. Grav. 12, 8 (1995): 1977

\bibitem{maart} Maartens R., Class. Quant. Grav. 12, 6 (1995): 1455

\bibitem{avelino} Avelino A. {\it et al.}, JCAL 8 (2013): 12.

\bibitem{novello} Novello M. and d'Olival J. B. S., Acta Phys. Pol. B 11 (1980): 3

\bibitem{jou}  Jou D., Extended Irreversible Thermodynamics, ed. Springer (1996)

\bibitem{mm} Maartens R. and M\'endez V., Phys. Rev. D 55, 4 (1997) 1937

\bibitem{chim} Chimento L. P. {\it et al.}, Class. Quantum Grav. 14 (1997): 3363-3375

\bibitem{ama} Amanullah R. {\it et al.}, A. J. 716, 1 (2010): 712-738

\bibitem{bamba} Bamba K. {\it et al.}, Astrophys. Space Sci. 342, 1 (2012): 155-228

\end{thebibliography}
\end{document}